\documentstyle[12pt,psfig,draft,lscape,array,amssymb]{nature-ejb-new}

\def\simgt{\stackrel{>}{{}_\sim}}
\def\simlt{\stackrel{<}{{}_\sim}}

\begin{document}

\title{\Large\bf
       The $\gamma$-ray burst GRB060614 requires a novel explosive process
       }

\author{\small
   A.~Gal-Yam,\affiliation[1]
     {\scriptsize Division of Physics, Mathematics and Astronomy, California Institute of 
     Technology, Pasadena, CA\,91125, USA \vspace{0.1in}}
   D.~B.~Fox,\affiliation[2]
     {\scriptsize Department of Astronomy and Astrophysics,
Pennsylvania State University, 525 Davey Lab, University Park, PA 16802, USA \vspace{0.1in}}
   P.~A.~Price,\affiliation[3]
     {\scriptsize University of Hawaii, Institute of Astronomy, 2680 Woodlawn Drive, Honolulu, 
       HI 96822-1897, USA \vspace{0.1in}}
   M.~R.~Davis,\affiliation[4]
     {\scriptsize Department of Astronomy, San Diego State University, San Diego,
California 92182, USA \vspace{0.1in}}
   D.~C.~Leonard,\affiliationmark[4]
   A.~M.~Soderberg,\affiliationmark[1]
   E.~Nakar,\affiliationmark[1]
   E.~O.~Ofek,\affiliationmark[1]
   B.~P.~Schmidt,\affiliation[5]
   {\scriptsize Research School of Astronomy and Astrophysics, Australian National University, Mt Stromlo Observatory,
via Cotter Rd, Weston Creek, ACT 2611, Australia \vspace{0.1in}}
   K.~Lewis,\affiliationmark[5]
   B.~A.~Peterson,\affiliationmark[5]
   S.~R.~Kulkarni,\affiliationmark[1]
   E.~Berger,\affiliation[6]
   {\scriptsize Observatories of the Carnegie Institution of Washington, 813 Santa Barbara Street, Pasadena, CA 91101, and
   Princeton University Observatory, Peyton Hall, Ivy Lane, Princeton, NJ 08544, USA \vspace{0.1in}}
   S.~B.~Cenko,\affiliationmark[1]
   R.~Sari,\affiliationmark[1]
   K.~Sharon,\affiliation[7]
   {\scriptsize Department of Astrophysics, Tel Aviv University, 69978 Tel Aviv, Israel \vspace{0.1in}}
   D.~Frail,\affiliation[8]
   {\scriptsize National Radio Astronomy Observatory, P.O. Box 0, Socorro, New Mexico 87801, USA \vspace{0.1in}}
   N.~Gehrels,\affiliation[9]
   {\scriptsize NASA/Goddard Space Flight Center, Greenbelt, Maryland 20771, USA \vspace{0.1in}}
   J.~A.~Nousek,\affiliationmark[2]
   D.~N.~Burrows,\affiliationmark[2]
   V.~Mangano,\affiliation[10]
   {\scriptsize INAF, Istituto di Astrofisica Spaziale e Fisica Cosmica di Palermo,
   Via Ugo La Malfa 153, I-90146, Palermo, Italy \vspace{0.1in}}
   S.~T.~Holland,\affiliationmark[9]
   P.~J.~Brown,\affiliationmark[2]
   D.-S.~Moon,\affiliationmark[1]
   F.~Harrison,\affiliationmark[1]
   T. Piran\affiliation[10]
   {\scriptsize Racah Institute of Physics, Hebrew University, Jerusalem 91904, Israel \vspace{0.1in}}
   S.~E.~Persson,\affiliationmark[6]
   P.~J.~McCarthy,\affiliationmark[6]
   B.~E.~Penprase,\affiliation[11]
   {\scriptsize Pomona College Dept. of Physics \& Astronomy, 610 N. College Ave, Claremont, CA 91711, USA \vspace{0.1in}}
   \& R.~A.~Chevalier\affiliation[12]
     {\scriptsize Department of Astronomy, University of Virginia, PO Box 3818, Charlottesville, VA 22903, USA \vspace{0.1in}}
}

\date{\today}{}
\headertitle{A new explosive origin for GRBs}
\mainauthor{Gal-Yam et al.}

\summary{\small%

Over the past decade our physical understanding of gamma-ray bursts  
(GRBs) has progressed rapidly thanks to the discovery and observation  
of their long-lived afterglow emission. Long-duration  
($T\simgt 2$ s) GRBs are associated with the explosive deaths of  
massive stars (``collapsars''\cite{MWH01}), which produce accompanying  
supernovae\cite{GVV+98,SMG+03,HSM+03}, 
while the short-duration ($T\simlt 2$ s) GRBs arise from a different origin, which has 
been argued to be the merger of two compact objects\cite{GSO+05,BPP+06,BPC+05},  
either neutron stars or black holes. Here we present observations of  
GRB\,060614, a 100-s long burst discovered by the Swift satellite\cite{PCG+06}, 
which require the invocation of a new explosive process:   
either a massive ``collapsar'' that powers a GRB without any associated  
supernova, or a new type of engine, as long-lived as the collapsar  
but without any such massive stellar host.  We also discuss the  
properties of this burst's redshift $z=0.125$ host galaxy, which  
distinguish it from other long-duration GRBs and suggest that an  
entirely new type of GRB progenitor may be required.
}

\maketitle



\begin{figure}
\centerline{\psfig{file=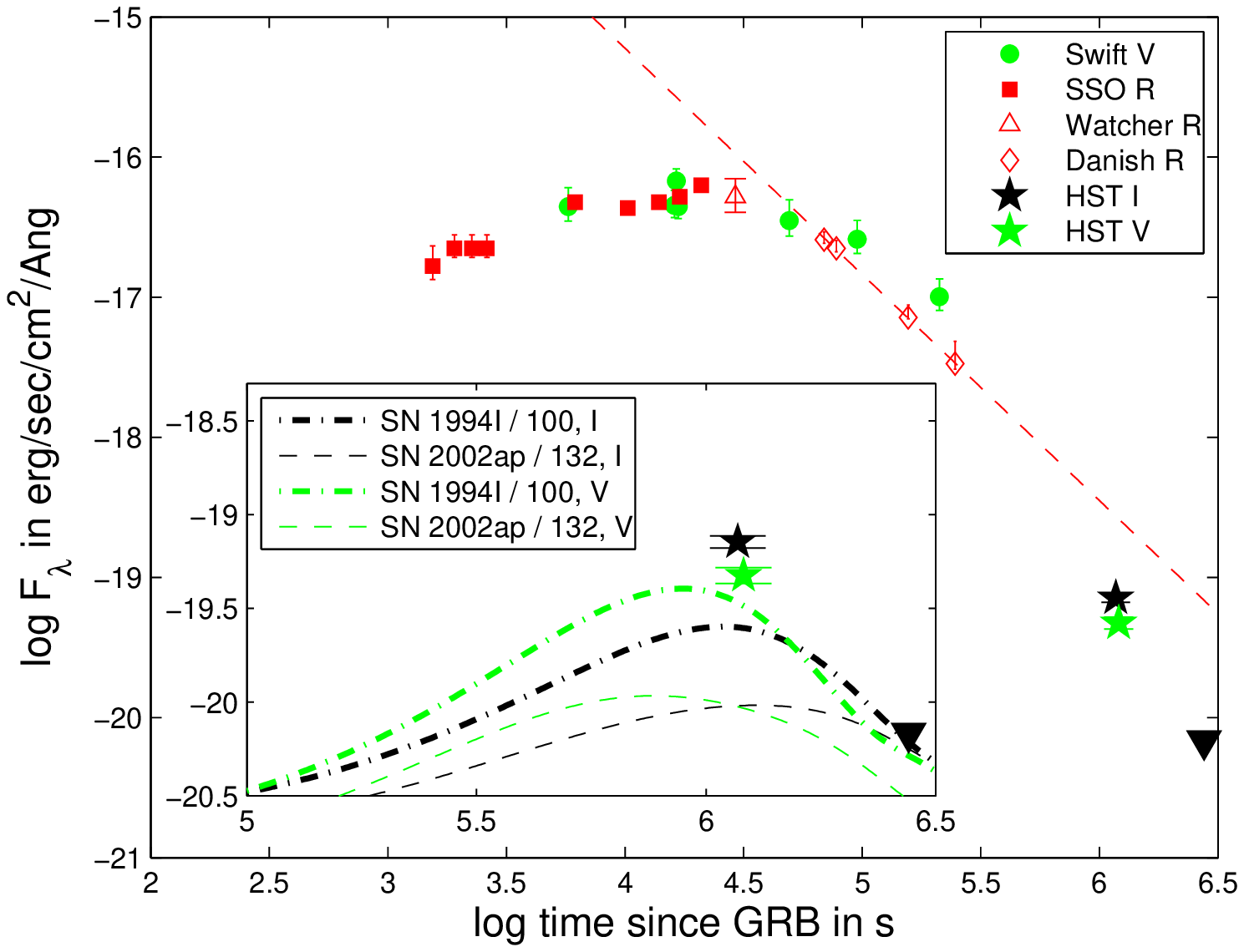,width=6.5in,angle=0}}
\bigskip
\caption[]{~\label{figlcs}~}
\end{figure}

\begin{figure}
\centerline{\psfig{file=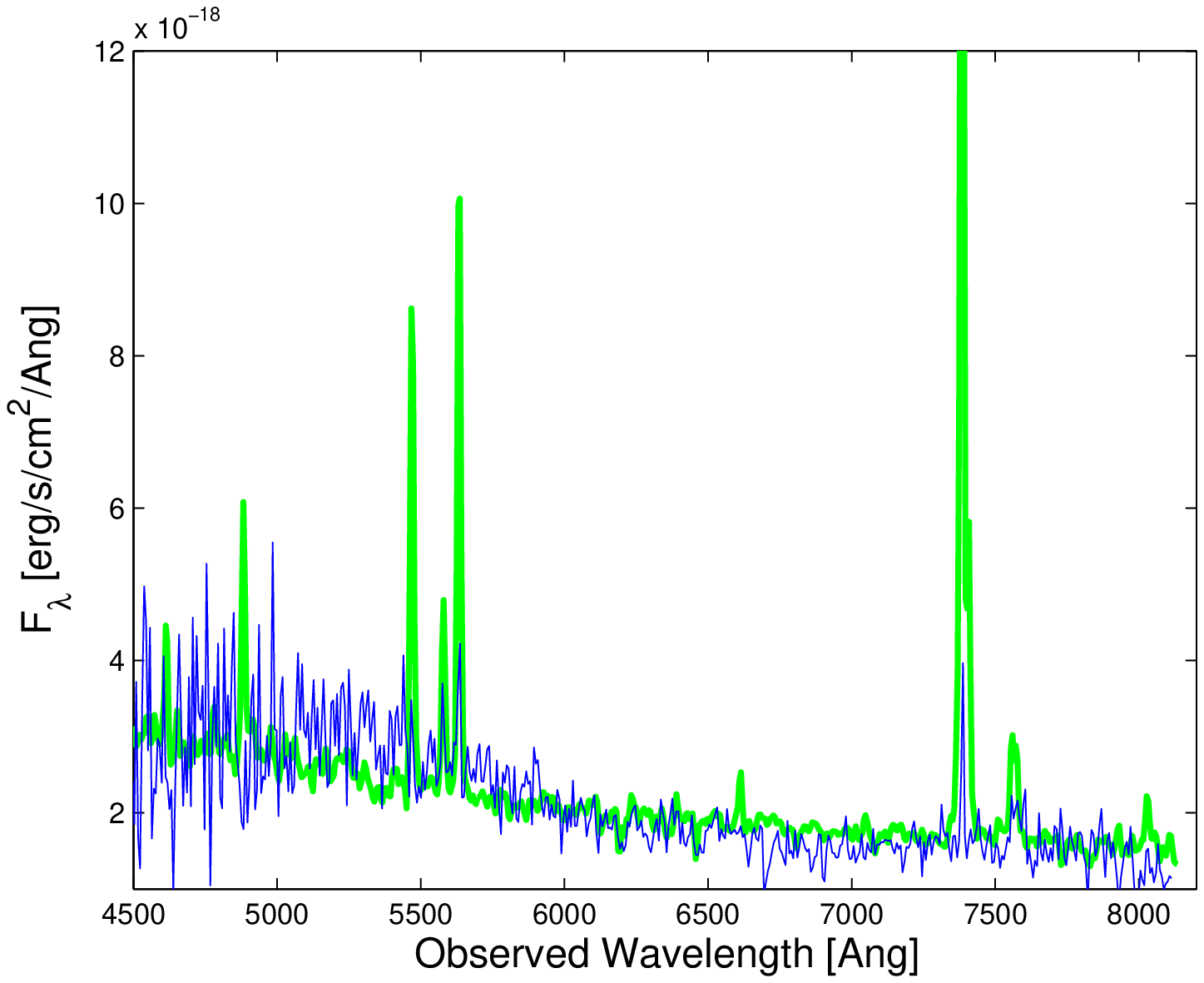,width=6.5in,angle=0}}
\bigskip
\caption[]{~\label{figspec}~}
\end{figure}

On 14 June 2006, 12:43 UT, the burst alert telescope (BAT) on board the {\it Swift} 
satellite detected the $\gamma$-Ray Burst (GRB) 060614\cite{PCG+06}. The BAT detected $\gamma$-rays
from this event for 120s, and the burst showed strong variability during much of that
period, as confirmed by parallel observations by the Konus-Wind 
satellite\cite{GAM+06}. Note that while some evolution in the temporal and spectral
properties of this GRB were observed, the emission remained highly variable and relatively
hard for tens of seconds, unlike the situation observed for a few short bursts with
long, soft ``tails''\cite{FFP+05,BPC+05}. This indicates sustained activity of an engine,
rather than the early onset of the afterglow.
The $\gamma$-ray properties of this event are similar to those of other bursts 
from the long-duration subgroup of GRBs. 
{\it Swift} autonomously slewed to the GRB position and began taking data with the X-ray telescope
and UV-optical telescope\cite{MLT+06}. We began observing this event $\approx26$
minutes later using the 40 inch telescope at Siding Springs Observatory. 
The evolution of the optical radiation from this event as traced by our
data, augmented by {\it Swift} observations and additional data from the literature 
is shown in Fig.~\ref{figlcs}. As the optical source decayed, we noticed
that it was apparently superposed on a faint dwarf host galaxy. On June 19, 2006
UT We obtained a spectrum of the host using the GMOS-S spectrograph mounted
on the Gemini-south 8m telescope at Cerro Pachon, Chile. From this spectrum 
we derived the redshift of the host galaxy, and by association of the GRB,
and found it to be $z=0.125$, a low value for long GRBs. We confirmed this 
redshift with a higher quality spectrum obtained using the same instrument 
on July 15, 2006 UT (Fig.~\ref{figspec}). Previous 
long GRBs at such low redshifts showed clear signatures of the underlying 
supernova (SN) explosions at comparable age
post-burst\cite{SMG+03,PMM+06}. However, such signatures were lacking in
the case of this long GRB\cite{FTJ+06}. 

Thus motivated, we undertook target-of-opportunity observations
with the {\it Hubble Space Telescope} (HST), combining resources from our
approved programs (GO 10551, PI Kulkarni, GO 10624, 10917, PI Fox). We
observed the location of GRB 060614 using the Wide Field and Planetary 
Camera 2 (WFPC2) on board HST on June 27-28, 2006 UT, and again using
the Advanced Camera for Surveys (ACS) on July 15-16 2006 UT. Inspection 
of the data (Fig.~\ref{fighst}) reveals a point source at the outskirts 
of the GRB host which is well-detected in our first-epoch WFPC2 observations,
and is apparently gone during our next visit. We identify this object as the optical
afterglow of GRB 060614, and derive its brightness using image-subtraction
methods. A decomposition of the measured flux into the contribution 
from the GRB afterglow (residual decaying radiation from the interaction 
of the GRB ejecta with itself and/or the surrounding material) and that
of a possible supernova (whose optical radiation is dominated by 
energy released from radioactive decay of newly synthesized elements,
mostly Ni 56), requires detailed modelling of the multiband evolution of
the GRB afterglow which is beyond the scope of this paper. However,
our analysis (Fig.~\ref{figlcs}) shows that our HST detection is 
probably dominated by the afterglow, rather than a possible SN, which
is not required by the data. Any putative SN component must be more than $100$
times fainter than the faintest event previously known to be associated
with a long GRB (SN 2006aj/ GRB 060218\cite{PMM+06,MSG+06}; Fig.~\ref{figlcs}). 
In fact, such a supernova would be fainter than any SN ever observed\cite{PZT+04}.  
A conservative upper
limit on the amount of synthesized Ni 56 is $8 \times 10^{-4}$ M$_\odot$, more than
two orders of magnitude less than the typical amount synthesized by long GRB/SNe. 
Our HST data thus indicate that this GRB was not associated with a radioactively-powered
event similar to any known SN.
   
Furthermore, our HST and ground-based data reveal that the properties of 
the host of GRB 060614 are unusual when compared to those of the large
sample of previously observed long GRBs. The star formation rate we measure
from the spectrum of the host, $0.0035$ M$_\odot$ y$^{-1}$, is very small, and 
even the specific star formation rate, correcting for the low luminosity 
of this dwarf galaxy ($L\approx0.015 L_*$) is 
$\sim0.23$ M$_\odot$ y$^{-1}$ $(L/L_*)^{-1}$,
a value which is more than $20$ times below that of typical long 
GRBs\cite{CHG04} and more than two orders of magnitude below 
that of long GRBs detected in low-luminosity dwarfs at low redshift
\cite{GPS+05,PBC+04}. 

It is worthwhile to note in this context that unlike long GRBs, members
of the short GRB group have been shown to reside in host galaxies
of all types, including elliptical galaxies with virtually no young,
massive stars\cite{BPP+06,BPC+05,GSO+05}, and in the outskirts of dwarf
galaxies\cite{FFP+05}, indicating a mature (rather than short-lived) 
population of progenitors (and no associated SNe). 

Considering the entire set of observations available for this
event, the emerging picture is a puzzling one. On the one hand,
the high-energy ($\gamma$-ray) properties of this burst are only
consistent with those of a long GRB. On the other hand, the lack of 
an associated SN is inconsistent with an origin in a massive,
rapidly-rotating star undergoing a core-collapse SN explosion 
(``collapsar''\cite{MWH01}),
the popular, observationally supported model for long GRBs. 
Furthermore, the properties of the host galaxy of this event stand out
from among those of numerous other long GRBs observed so far,
and rather resemble those of a short GRB. 
We now consider possible explanations for this puzzle, and their implications. 

First, we can conjecture that all long GRBs result from massive ``collapsars'',  
but that most of these are associated with SNe with a range of properties
(as observed so far\cite{ZKH04,SKP+06}) and a minority (the first example of which is
GRB 060614) synthesize very little Ni 56, and do not produce an optically
luminous SN. Indeed, there are models of massive stars that collapse
into a black hole and produce, if at all, very faint SNe\cite{NTU+05}.
However, such events are understood essentially as non-rotating stars,
collapsing directly into a black hole,  while ``collapsar''-like GRB 
models generically require a rapidly rotating core, leading to an accretion
disk onto a spinning BH that powers and/or collimates the outgoing beamed
ultra-relativistic ejected material (but see\cite{M03}). 
The large angular momentum in such systems
requires an accretion disk to facilitate infall into the central BH, and
is thus inconsistent with the non-rotating, direct-infall models for
non-SN massive stellar collapses. In particular, GRB 060614 emitted 
a large flux ($10^{50}$ erg s$^{-1}$) of high-energy photons and displayed rapid variability,
indicating an ultra-relativistic flow with Lorentz factor of $\Gamma>15$
(calculated following the methodology presented by Ref. [23]).   
Such a flow is not expected in a nearly-spherical direct collapse of a star
into a BH, and we thus consider this option to be apparently
at odds with theoretical understanding of massive stellar death.

Second, especially given the remarkable properties of the host galaxy of
GRB 060614, one might suggest that this is an extreme member of the 
short GRB group, which are not associated with massive
stellar progenitors, and may result from compact
binary mergers. However, to maintain the putative association with
binary mergers would involve a major
revision in our understanding of accretion physics. In particular,
low viscosity during compact binary mergers are required to allow 
such a process to continue beyond a fraction of a second, 
in contrast with the current consensus\cite{NPK01}. Alternative
interpretation would be that this is indeed an extreme member
of the short GRB group, and that its existence indicates that this
entire group does not result from compact binary mergers (which can
only produce short duration bursts) but rather from a different
physical process altogether, capable of producing both short and
long events, and not associated with young massive stars.   
An obvious and testable prediction of these options is that such a ``long'' 
member of the ``short'' (binary merger or other mechanism) group should be detected in an
elliptical galaxy containing only old stars. Such galaxies host
a significant fraction of the ``short'' GRB population, but have never
so far been associated with events longer than a few seconds. 

Finally, GRB 060614 may be the first example of a new class of GRBs,
different from both typical long events (which are associated with SNe
and powered by infall onto a newly formed BH) and short events (which 
may come from compact binary mergers). Many additional mechanisms
have been suggested over the years to produce $\gamma$-ray bursts,
and GRB 060614 may be the first clear example of one of these alternative scenarios.
It is surprising, however, that three distinct processes (collapsars, binary
mergers, and ``the third mechanism'') would results in events which are
so similar in their high-energy properties. A possible explanation would be
that these are three different routes that begin with very different physical
systems but lead to a similar outcome, e.g., rapid accretion onto a BH. The
many similarities between short and long GRBs\cite{N06} may hint at this 
direction.

Regardless of the ultimate
resolution of this puzzle (following studies of GRB 060614 and possible
additional future examples of this class) it is already obvious that the
elegant simple picture, consisting of two groups of GRBs with distinct 
physical origins (long GRBs from SN/collapsars and short GRBs from 
binary mergers) which was briefly consistent with GRB observations and 
theory, must now be revised.    
\footnote{Correspondence should be addressed to A. Gal-Yam
(avishay@astro.caltech.edu).}


\begin{acknowledge}
A.G. acknowledges support by NASA through Hubble Fellowship grant
\#HST-HF-01158.01-A awarded by STScI, which is operated by AURA, Inc.,
for NASA, under contract NAS 5-26555.   
SRK's research is supported by NSF and NASA.
\end{acknowledge}


\bibliographystyle{nature-pap}
\bibliography{journals,refsGRB}

\clearpage

\noindent
{\bf Figure 1}: 
{%
Temporal evolution of the optical transient associated with 
GRB 060614. The main figure shows V-band observations from 
UVOT (solid circles), our SSO R-band data (solid squares), 
augmented by by data from GCN 5257\cite{FMH+06} (empty triangles)
and GCN 5277\cite{FTJ+06} (empty diamonds), along with our late time HST V- and I-band detections
(stars) and upper limit (Solid inverted triangle; see Fig.~\ref{fighst}). 
Note that our HST detections fall below 
projections based on early V-band data 
or the fit proposed\cite{FTJ+06} based on R-band data (red dashed curve), indicating
a steepening of the temporal decay prior to our first HST visit. The contribution
of the host galaxy, estimated as detailed in Fig.~\ref{figspec}, was removed from the
photometry.
The inset shows a comparison of our HST detections and upper limit
with scaled-down light curves of SNe, properly K-corrected and 
time-dilated\cite{SKP+06}. The brightest allowed SN is
obtained by assuming the steeply declining light curve of SN 1994I (heavy dash-dot)
and the maximal amount of extinction allowed by the analysis of
early UVOT data (see supplementary Fig.~\ref{figuvotspec}). In this
case the absolute magnitude of the SN will be $M_V=-12.6$, fainter
than any SN ever detected in the nearby Universe, synthesizing 
only $\sim8\times10^{-4}$ M$_\odot$ of Ni 56. More likely scenarios
involving moderate extinction values ($A_V=0.5$) or SNe with 
more slowly-decaying light curves such as SN 2002ap (thin dashed line), similar to 
the faintest GRB-associated SN 2006aj, would impose more stringent
limits on the luminosity and Ni 56 production of a putative SN 
(by factors of $\sim3$). Note that in all cases the emission
detected during our first HST visit must be dominated by the GRB
afterglow emission, with but a fraction of the light coming from
the peaking faint SN, and that the data are well-explained without
invoking any SN-like component, with the optical afterglow roughly following
the late X-ray decline rate\cite{MLT+06} (to which a SN is not expected to contribute).  
}

\clearpage

\noindent
{\bf Figure 2}: 
{%
A spectrum of the host galaxy of GRB 060614 (thin blue curve) obtained with the GMOS-S
spectrograph mounted on the Gemini-South 8m telescope at Cerro Pachon,
Chile, on July 15, 2006 UT. Four exposures of 1200s each were
reduced and combined in the usual manner within IRAF, including wavelength- 
and flux-calibration using the smooth spectrum standard star EG131.
The spectral continuum is similar to a template Sc galaxy spectrum (heavy green curve)
shifted to $z=0.125$ and scaled in flux. We note that the emission 
lines are much weaker, though. From the luminosity of the H$\alpha$ line we
derive an upper limit on the star-formation rate in the galaxy,
SFR = 0.0035 M$_\odot$ yr$^{-1}$. The ratio of H$\alpha$ to H$\beta$ line
strengths indicates that the emission-line regions in this
dwarf galaxy suffer negligible extinction. Assuming that the galaxy spectrum
is well-described by the Sc template throughout the optical range, we 
use synthetic photometry anchored to the V-band magnitude of the galaxy
measured from our HST images to determine the magnitudes of the host 
to be [UBVRI]=[23.24 23.89 23.66 23.47 22.87] mag, respectively. The host
contribution is removed from the photometry presented in Fig.~\ref{figlcs}. 
}

\clearpage

\noindent
{\bf Figure 3}: 
{%
HST observations of the location of GRB 060614. Upper panels show
I-band (F814W filter) data and lower panels V-band (F606W) data. 
Images were obtained using the WFPC2 camera (left, 6000s total exposure
time; F814W filter; UT dates as marked) and the ACS camera (middle panels;
total exposure 3600s; same filter). The third panels show 
the difference between the first epoch images and a third epoch visit 
with ACS (I-band; July 29, 2006 UT, total exposure time 4840s) or the second
epoch in the case of the V-band, calculated using our image subtraction
methods\cite{GLF+06}. WFPC2 data were reduced and photometered using 
an adaptation of HSTphot\cite{GLF+06} and ACS data were reduced in the 
standard manner with IRAF/multidrizzle. We have used HSTphot-calibrated, 
nearby, isolated, compact sources to establish a calibration grid of
Johnson-V and Cousins-I local standards and photometered the afterglow     
with respect to this grid using the image-subtraction-based photometry
pipeline mkdifflc\cite{GLF+06}. A similar comparison between the second and
third I-band HST visits (July 16 and 29 UT, respectively) shows no residual
to a 4$\sigma$ upper limit of I=28.1 mag, indicating that the optical 
transient was undetectable during our second visit and justifying the use
of the second-epoch V-band image as a subtraction template. 
The resulting photometry and upper limit are reported in Fig.~\ref{figlcs}. 
Note the overall regular structure of this faint dwarf host and the 
peripheral location of the optical transient.   
The orientation of the images 
is marked with a long arrow due north and a short arrow
due east. The length of the long arrow is $2.5''$ for 
scale.
}

\begin{figure}
\bigskip
\caption[]{~\label{fighst}~}
\label{fighst}
\end{figure}

\clearpage

\noindent
{\bf Supplementary Figure 4}: 
{%
A photometric spectrum derived from early multi-band
UVOT observations. The flux in the 5 broad-band UVOT 
filters (V,B,U,UVW1,UVW2 with effective wavelengths
of 550, 450, 330, 250 and 180 nm, respectively) was
interpolated to 10000s after the GRB. We consider
a hot thermal spectrum ($F_{\nu} \propto \nu^{2}$) to be
the bluest possible intrinsic spectrum at this time
(such a blue spectrum has never been observed in any 
GRB). The observed spectrum cannot be dereddened by
more than $A_V=1.7$ without becoming bluer than this
hot spectrum (red curve and tiangles). A more typical
blue intrinsic spectrum for a GRB afterglow with 
$F_{\nu} \propto \nu^{0.5}$ allows only moderate
extinction ($A_V=0.5$). Fits of similar quality 
are also obtained without invoking any extinction. 
At later times (up to $10^5$ s) the observed photometric
spectrum becomes redder, imposing weaker constraints
on the extinction. We adopt $A_V=1.7$ as a conservative
upper limit on the extinction toward this optical transient,
but estimate that the more likely value is actually
lower ($A_V\approx0.5$ or less). 
}

\begin{figure}
\centerline{\psfig{file=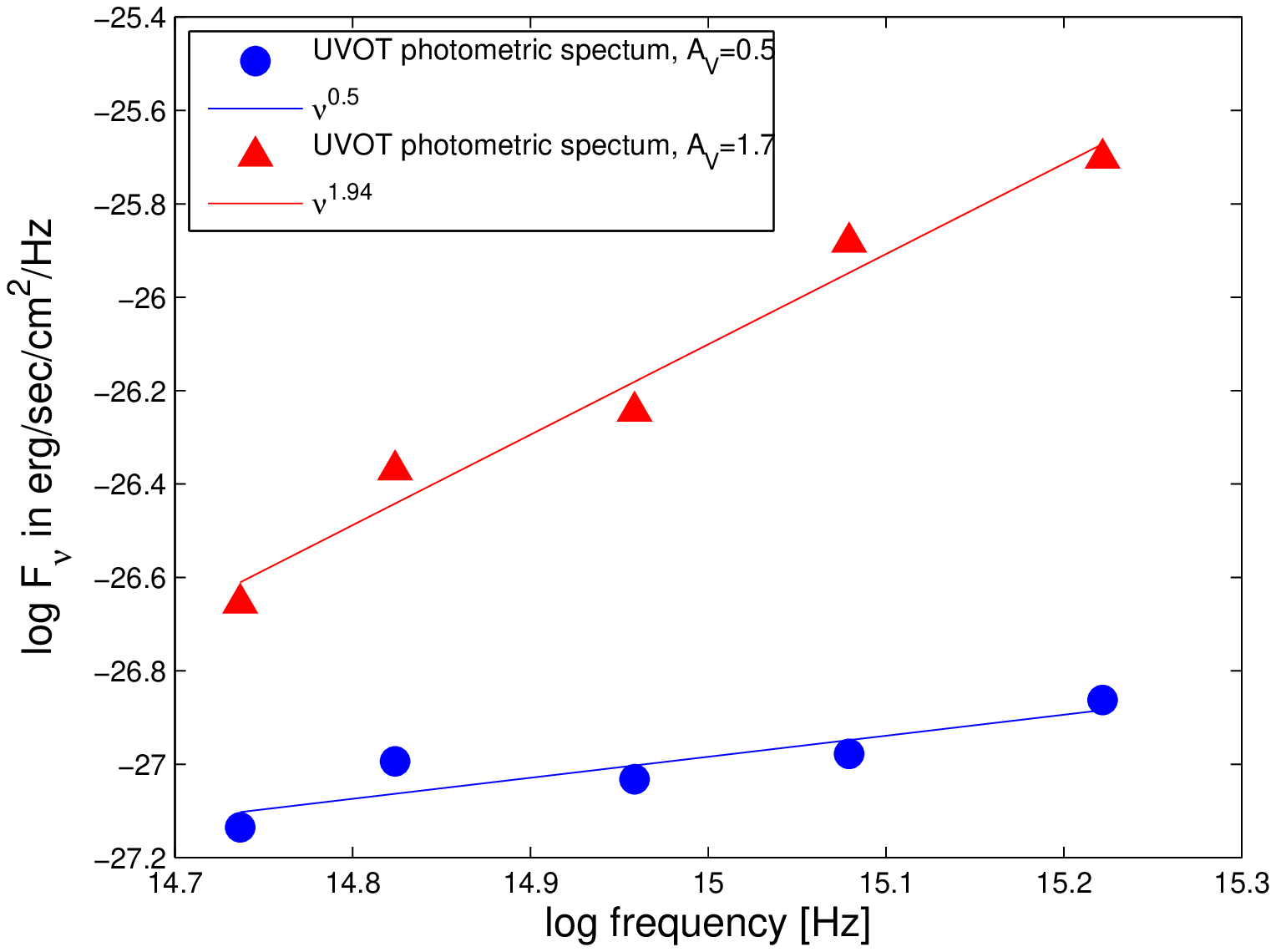,width=6.3in,angle=0}}
\bigskip
\caption[]{~\label{figuvotspec}~}
\end{figure}

\clearpage

%
%
%
%

\end{document}